\documentclass[aps,prl,twocolumn]{revtex4}%
\usepackage{amsfonts}
\usepackage{amsmath}
\usepackage{amssymb}
\usepackage{graphicx}%
\providecommand{\U}[1]{\protect\rule{.1in}{.1in}}

\begin{document}
\title{Pairing dynamics in strongly correlated superconductivity}
\author{B. Kyung, D. S\'{e}n\'{e}chal, and A.-M.S. Tremblay}
\affiliation{D\'{e}partement de physique and Regroupement qu\'{e}b\'{e}cois sur les
mat\'{e}riaux de pointe, Universit\'{e} de Sherbrooke, Sherbrooke, Qu\'{e}bec
J1K 2R1, Canada.}

\begin{abstract}
Confirmation of the phononic origin of Cooper pair formation in
superconductors came with the demonstration that the interaction was retarded
and that the corresponding energy scales were associated with phonons. Using
cellular dynamical mean-field theory for the two-dimensional Hubbard model, we
identify such retardation effects in d-wave pairing and associate the
corresponding energy scales with short-range spin fluctuations. We find which
frequencies are relevant for pairing as a function of interaction strength and
doping and show that the disappearance of superconductivity on the overdoped
side coincides with the disappearance of the low energy feature in the
antiferromagnetic fluctuations, as observed in neutron scattering experiments.

\end{abstract}
\date{\today}

\maketitle

In ordinary superconductors, the origin of attraction between electrons, the
\textquotedblleft pairing glue\textquotedblright, manifests itself in
observable quantities. Indeed, the characteristic frequencies of phonons
appear directly in the frequency dependence of the gap function, which in turn
enters observables such as the single-particle density of states or the
infrared conductivity. Migdal-Eliashberg theory
\cite{McMillan:1965,Carbotte:1990} has been extremely successful to extract
from these experiments the spectral function of the phonons that provide the glue.

High-temperature superconductors, heavy-fermion and layered organic
superconductors all have phase diagrams where non s-wave superconducting order
parameters lie in close proximity to antiferromagnetic phases. In the case of
high-temperature superconductors, much effort has been devoted to find out
whether antiferromagnetic fluctuations could be the pairing
glue\ \cite{Bickers_dwave:1989,Carbotte:1999,Chubukov:2003,Monthoux:2007,Maier:2008}%
. Even though its assumptions are not generally valid in that case, Eliashberg
theory has been used to extract the amplitude and frequency dependence of a
spectral function that is found to be similar to that for antiferromagnetic
fluctuations directly measured by neutron scattering
\cite{Zasadzinski:2006,Hwang:2008,Hwang:2008b,vanHeumen:2008}.

But understanding the origin of pairing in high-temperature superconductors
requires an approach that does \textit{not} rely on the assumptions entering
Eliashberg theory and that takes into account Mott insulating behavior. This
seems to rule out theories that are based purely on early weak-coupling ideas
of boson exchange \cite{Beal-Monod:1986,Scalapino:1986,Miyake:1986}. In fact,
Anderson \cite{Anderson:2007} has argued that the appropriate starting point
consistent with Mott physics is the strong-coupling version of the Hubbard
model, or the $t-J$ model, and that in this case interactions are
instantaneous, as suggested by mean-field factorization \cite{Kotliar:1988}.

In this paper, we show, for the Hubbard model, that spectral features of the
imaginary part of the anomalous (off-diagonal) self-energy do correspond to
those of the spectral function for short-range spin fluctuations and that the
energy scales relevant for pairing are also those of spin fluctuations.

The Hubbard model Hamiltonian is given by
\begin{equation}
H=-\sum_{i,j,\sigma}t_{ij}c_{i,\sigma}^{\dagger}c_{j,\sigma}+U\sum
_{i}n_{i\uparrow}n_{i\downarrow}%
\end{equation}
where $t_{ij}$ and $U$ correspond to the hopping matrix and the onsite
screened Coulomb repulsion respectively with $c_{i,\sigma}^{(\dagger)}$ the
destruction (creation) operators for an electron at site $i$ with spin
$\sigma$ and $n_{i\sigma}=c_{i,\sigma}^{\dagger}c_{i,\sigma}$ the number
operator. The theoretical method that has been most successful to date to
treat the Mott transition starting from the one-band Hubbard model is
dynamical mean-field theory (DMFT) \cite{Georges:1996}.

Cluster generalizations of DMFT
\cite{Hettler:1998,Kotliar:2001,Potthoff:2003b,Maier:2005} are necessary to
study problems in two dimensions where correlations beyond single site must be
taken into account to study, for example, d-wave superconductivity. They lead
to phase diagrams that have the same features as those observed experimentally
for both electron- and hole- doped high-temperature superconductors
\cite{Senechal:2005,Kancharla:2008,Haule:2007,Macridin:2005} and for organic
conductors. In addition, observable quantities such as the density of states,
\cite{Haule:2007} the ARPES spectrum \cite{Kancharla:2008,Haule:2007} and the
optical conductivity \cite{Haule:2007} have the experimentally observed
behavior. The method that we use, C-DMFT with exact diagonalization at $T=0$,
is described in Refs.\ \cite{Kotliar:2001,Kancharla:2008} and in
Ref.\ \cite{SupplementRetardation:2009}. We stress that it does not involve
\textit{any} Eliashberg-like approximation.

In C-DMFT, antiferromagnetism and d-wave superconductivity coexist over part
of the phase diagram. This is seen in stoichiometric cuprates with
intrinsically doped planes \cite{Mukuda:2008} and in a few other
cases,\ \cite{Dai:2005,Chang:2007} but does not appear to be a completely
generic property of the phase diagram. Our work is thus restricted to showing
that spin fluctuations are relevant for high-temperature
superconductivity all the way to the overdoped regime, leaving open the
possibility that additional types of fluctuations may either contribute to or
hinder superconductivity in the underdoped phase.

\begin{figure}[ptb]
\includegraphics[width=8cm]{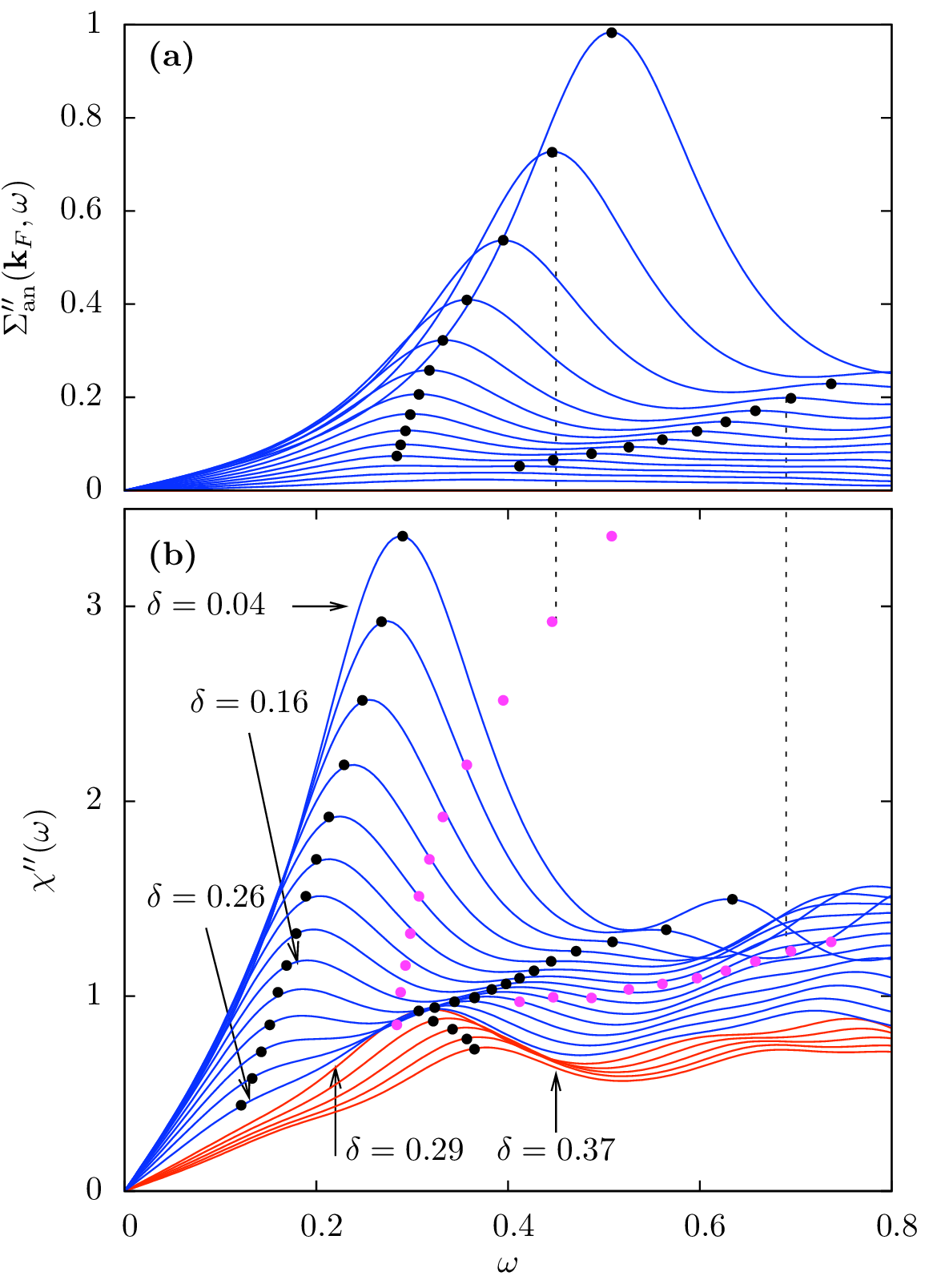}\caption{(Color
online) (a) Imaginary part of the anomalous self-energy $\operatorname{Im}%
\Sigma_{an}\equiv\Sigma_{an}^{\prime\prime}$ at the Fermi wave vector nearest
to the antinodal point, for various dopings. In (b) Imaginary part of the
local spin susceptibility $\operatorname{Im}\chi\equiv\chi^{\prime\prime}$.
Black dots in (a) and (b) identify peaks. The position of the peaks of
$\Sigma_{an}^{\prime\prime}$ in (a) are reported as magenta dots in (b) at the
same height as the corresponding $\chi^{\prime\prime}$ to illustrate the
correspondence between the main peaks of the two functions. The frequency
splitting between the peaks decreases with doping, like the single particle
gap. The red curves are for the normal state. The lower frequency peak present
in the superconducting state disappears and the next peak moves to higher
frequency with doping. In all the figures Lorentzian broadening is $0.125t,$
$U=8t,$ $t^{\prime}=-0.3t,$ $t^{\prime\prime}=-0.08t$, $t=1,\hbar=1$, for
La$_{2-x}$Sr$_{x}$CuO$_{4}.$}%
\label{Fig_ImSigma_Im_Chi}%
\end{figure}

The correspondence between the imaginary part of the anomalous self-energy,
$\Sigma_{an}^{\prime\prime},$ and the imaginary part of the local spin
susceptibility, $\chi^{\prime\prime}$, is seen in
Fig.\ \ref{Fig_ImSigma_Im_Chi}. We take band parameters appropriate for
La$_{2-x}$Sr$_{x}$CuO$_{4},$ namely $t^{\prime}=-0.17t$ for nearest-neighbor
and $t^{\prime\prime}=0.08t$ for next-nearest-neighbor hopping. CDMFT with
$U=8t$ then leads to superconductivity in the doping range observed
experimentally \cite{Kancharla:2008}. The anomalous self-energy $\Sigma
_{an}^{\prime\prime}$ is defined as minus the off-diagonal part of the inverse
Green function in Nambu space. Numerical results are presented in energy units
where $t=1.$
For all different dopings, the positions of the first two peaks in the spin
fluctuations (black dots on middle panel) are just shifted down with respect
to the corresponding peaks in $\Sigma_{an}^{\prime\prime}$ (black dots on top panel).

In Eliashberg theory for the electron-phonon interaction, the first two peaks
in the phonon density of states are shifted down with respect to those in
$\Sigma_{an}^{\prime\prime}$ by the BCS gap \cite{Maier:2008}. Similarly, the
down shift of peaks in $\chi^{\prime\prime}$ seen in Fig.
\ref{Fig_ImSigma_Im_Chi}b increases as we underdope, like the single-particle
gap. For $U=12t$ and realistic band structure for YBa$_{2}$Cu$_{3}$O$_{7-x}$
the shift is very weakly doping dependent \cite{SupplementRetardation:2009}.

In Migdal-Eliashberg theory, the real part of the self energy $\Sigma
_{an}^{\prime}$ times the quasiparticle renormalization factor is the gap
function. We find that this function, has no static (frequency independent)
contribution,\ \cite{SupplementRetardation:2009} contrary to what was found in
the $t-J$ model.\ \cite{Haule:2007,Maier:2008}.

\begin{figure}[ptb]
\includegraphics[angle=0,width=8cm]{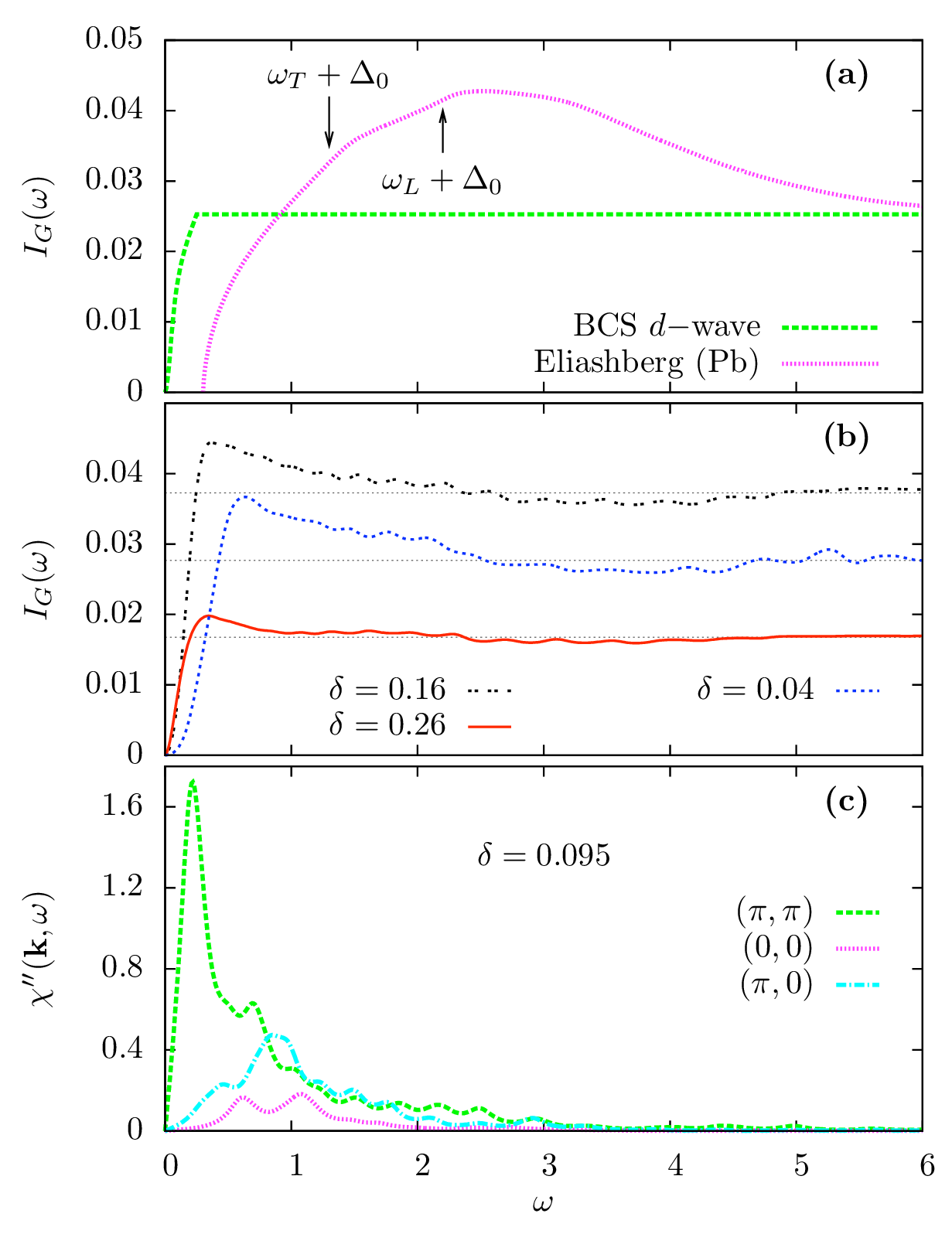}\caption{(Color online)
(a) The dashed green line is $I_{G}\left(  \omega\right)  $ for a d-wave BCS
superconductor with a cutoff at $\omega_{c}=0.5.$ The magenta line is obtained
from Eliashberg theory for Pb in Ref. \onlinecite{Scalapino:1966}. Frequencies
in that case are measured in units of the transverse phonon frequency. The two
glitches before the maximum correspond to the transverse and longitudinal
peaks in the phonon density of states. The scale of the vertical axis is
arbitrary. (b) $I_{G}\left(  \omega\right)  $ calculated for various dopings.
The horizontal lines for the asymptotes mark the value of the order parameter.
(c) The three independent Fourier components of $\chi^{\prime\prime}\;$on a
$2\times2$ plaquette for an underdoped case.The $\left(  \pi,\pi\right)  $
component dominates at low frequencies.}%
\label{Fig_IG_wide}%
\end{figure}

To identify the energy scales relevant for the pairs, we introduce
the function 
\begin{equation}
I_{G}\left(  \omega\right)  \equiv-\int_{0}^{\omega}\frac{d\omega^{\prime}%
}{\pi}\operatorname{Im}F_{ij}^{R}\left(  \omega^{\prime}\right)  .
\end{equation}
$F^{R}$ is the retarded Gork'ov function defined in imaginary time by
$F_{ij}\equiv-\langle Tc_{i\uparrow}(\tau)c_{j\downarrow}(0)\rangle$
with $i$ and $j$ nearest-neighbors. The infinite frequency limit of
$I_{G}\left(  \omega\right)  $ is equal to $\left\langle c_{i\uparrow
}c_{j\downarrow}\right\rangle $ which in turn is proportional to the $T=0$
d-wave order parameter (it changes sign under $\pi/2$ rotation). It was shown
in Ref. \cite{Haule:2007} that $\left\langle c_{i\uparrow}c_{j\downarrow
}\right\rangle $ scales like $T_{c}$. For all these reasons, $I_{G}\left(
\omega\right)  $ is useful to estimate the frequencies relevant for binding.
Its meaning is illustrated by the d-wave BCS result in Fig. \ref{Fig_IG_wide}%
a). The function $I_{G}\left(  \omega\right)  $ rises monotonically until it
reaches the sharp BCS cutoff frequency $\omega_{c}$ above which no binding
occurs. $I_{G}\left(  \omega\right)  $ extracted from the Eliashberg
calculation \cite{Scalapino:1966} for lead is also displayed in Fig.
\ref{Fig_IG_wide}a). The maximum is reached at a frequency just above the
largest phonon frequency.

$I_{G}\left(  \omega\right)  $ is plotted in Fig.\ \ref{Fig_IG_wide}b for
underdoping $\delta=0.4,$ optimal doping $\delta=0.16$ and overdoping
$\delta=0.26$. The asymptotic large frequency value of $I_{G}\left(
\omega\right)  $ indicated by horizontal lines gives the order parameter that,
as a function of doping, has the dome shape dependence \cite{Kancharla:2008}.
The functions $I_{G}\left(  \omega\right)  $ cross their respective asymptotic
values at progressively lower frequencies as doping increases. The spin
fluctuations that dominate at the lower frequencies come from wave vectors
around $\left(  \pi,\pi\right)  ,$ as illustrated in Fig.\ \ref{Fig_IG_wide}c
for an underdoped case. The maximum of $I_{G}\left(  \omega\right)  $ is more
pronounced in the underdoped regime. The form of $I_{G}\left(  \omega\right)
$ in the overdoped regime is closer to the BCS limit with just a sharp cutoff.
Our calculations are less precise at high frequencies, but nevertheless they
suggest that, in all cases, $I_{G}\left(  \omega\right)  $ undershoots very
slightly its asymptotic value and then recovers at frequencies that are of
order $U/2$ where the upper Hubbard band opens new scattering channels
\cite{Maier:2008}$.$ This has no analog in ordinary superconductors.

\begin{figure}[ptb]
\includegraphics[angle=-90,width=8cm]{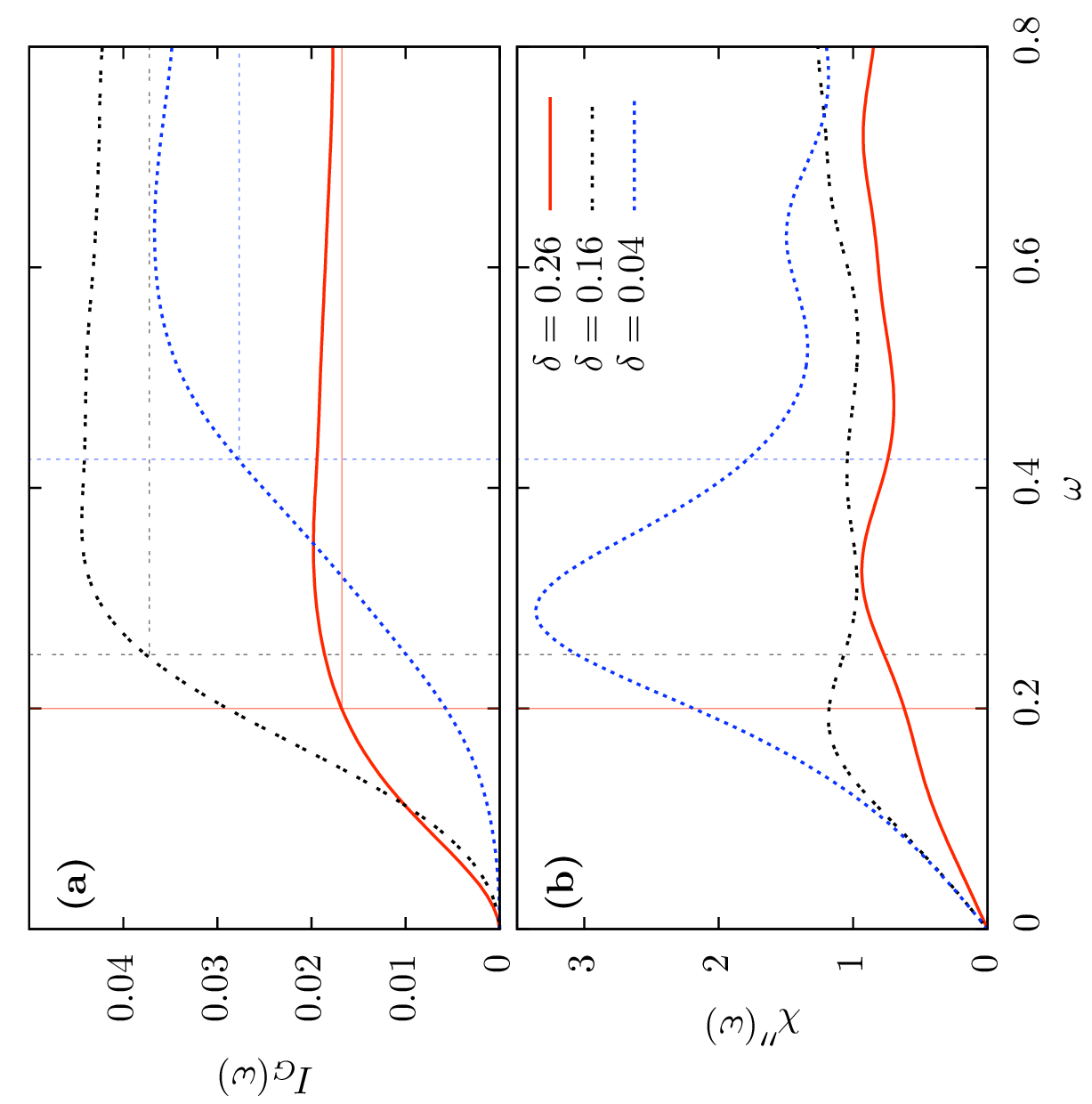}\caption{(Color online)
(a) Low frequency part of $I_{G}\left(  \omega\right)  $ for three dopings,
underdoped, optimally doped and overdoped. The vertical lines indicate the
first intersection with the asymptotic value. The plot of $\chi^{\prime\prime
}$ in (b) shows that the first peak occurs in the frequency interval where
$I_{G}\left(  \omega\right)  $ reaches its maximum. These are the main
frequencies that give rise to pair binding. }%
\label{Fig_IG_low}%
\end{figure}

In Fig. \ref{Fig_IG_low} we focus on the low-frequency behavior. On the top
panel, $I_{G}\left(  \omega\right)  $ crosses its asymptotic value for the
first time near its maximum. This crossing point shown by vertical lines
follows the first peak in the corresponding $\chi^{\prime\prime}$ in the
bottom panel. By studying the cases $U=8,12,16$ we have verified that these
features scale with $J$. Clearly, if we wished to design an approximate
mean-field theory\ \cite{Anderson:2004} for this problem that would play a
role analogous BSC theory as an approximation of the Migdal-Elisahberg theory,
we would use a frequency cutoff of order $J.$

Let us now discuss how the properties of the spin fluctuations $\chi
^{\prime\prime}\left(  \omega\right)  $ correlate with those of the d-wave
superconducting state for La$_{2-x}$Sr$_{x}$CuO$_{4}.$ In Fig.
\ref{Fig_ImSigma_Im_Chi}b, one sees that in the underdoped regime the low
frequency peak is the most prominent feature. Optimal doping $\left(
\delta\sim0.16\right)  $ is reached when the intensity of the low frequency
peak becomes comparable to the next one at higher frequency. More importantly
one sees that, around doping $\delta=0.26\,$, superconductivity disappears
with the low frequency peak in $\chi^{\prime\prime}\left(  \omega\right)  $
below $\omega$ about $J/2=0.25t$. That low frequency peak is the one involved
in the dynamics of the pairs as shown by the intersection of the
$I_{G}\left(  \omega\right)  $ function with its asymptotic value in Fig.
\ \ref{Fig_IG_wide}b. The leftover peak in the normal state, indicated in red
in Fig. \ref{Fig_ImSigma_Im_Chi}b, moves to higher frequencies as we dope.
These features are those found in neutron scattering
experiments.\ \cite{Wakimoto:2004,Wakimoto:2007}

The large tails and monotonic decrease of the weight of the low frequency peak
in $\chi^{\prime\prime}$ as we dope are consistent with the \textquotedblleft
glue function\textquotedblright\ extracted from recent optical conductivity
experiments.\ \cite{vanHeumen:2008}\ The position of the low frequency peak
near $0.2t$ at optimal doping is consistent with the experimental value
\cite{vanHeumen:2008} if we take $t=250\;\mathrm{meV}.$ One should recall that
Fig.\ \ref{Fig_ImSigma_Im_Chi}b for the local spin spectral weight
$\chi^{\prime\prime}$ gives information integrated in wave vector so the
properties of the \textquotedblleft neutron resonance\textquotedblright%
\ located at $\left(  \pi,\pi\right)  $ have to be found by a different
approach. In a recent calculation with a related cluster method,
\cite{Brehm:2008} it has been found that the peak located at $\left(  \pi
,\pi\right)  $ in the infinite lattice decreases with frequency in the
underdoped regime. Given the small weight of this \textquotedblleft neutron
resonance\textquotedblright, this does not contradict the fact that
$\chi^{\prime\prime},$ whether local or averaged over one-quarter of the
Brillouin zone near $\left(  \pi,\pi\right)  $, has the opposite doping
dependence \cite{Fong:2000, Lipscombe:2007, Vignolle:2007,
Wakimoto:2004,Wakimoto:2007} The magnetic resonance itself has small weight
\cite{Kee:2002}.

We stress that, despite the similarities, the results obtained in this paper
are not identical to those that are obtained in ordinary superconductors. In
particular, in the underdoped regime the spin fluctuations are strongly
pair-breaking. The pair-breaking effect of the pseudogap can be seen from the
fact that in the normal state, the pairing susceptibility decreases as one
approaches half-filling when vertex corrections are neglected
\cite{SupplementRetardation:2009}. Following the suggestion of Ref.
\cite{Schiro:2008}, we also checked whether the pseudogap is pair breaking by
computing the ratio $\Sigma_{an}\left(  i\omega_{n}\right)  /(1-\Sigma
_{n}\left(  i\omega_{n}\right)  /i\omega_{n})$ at the antinodal Fermi surface
crossing as a function of Matsubara frequency. The ratio is far from constant,
in agreement with the existence of strong pair-breaking effects in the
pseudogap. From a diagrammatic point of view, spin fluctuations both
scatter electrons (self-energy), decreasing the density of states at the Fermi
surface, and provide the glue (vertex). This can lead to the dome shape
of the transition temperature as a function of doping \cite{Kyung:2003}.
In the underdoped regime, the pair-breaking effect wins over
the glue provided by the vertex, whereas in the overdoped regime the vertex dominates.

In conclusion, we have found that the imaginary part of the anomalous
self-energy has a structure that is correlated with the spectrum of spin
fluctuations. This correlation is similar to the one found with the phonon
spectrum in the Migdal-Eliashberg theory of ordinary superconductors. This
suggests the importance of spin-one excitations in pair formation. Our approach also allows for mutual feedback between spin fluctuations and pairing \cite{Bourbonnais:2009}. The
frequencies most relevant for the pair
dynamics scale with the Heisenberg exchange $J$. Superconductivity
disappears for sufficient overdoping when the first peak in $\chi
^{\prime\prime}$ below frequencies about $J/2$ becomes negligible. There are,
however, major differences with ordinary superconductors. Even though the
anomalous self-energy increases as we approach half-filling, the order
parameter decreases because of large self-energy effects in the normal part of
the propagator. These come from Mott Physics at half-filling. The magnetic
fluctuations that we find have the same doping and energy dependence as that
found in optical,\ \cite{Hwang:2008,Hwang:2008b,vanHeumen:2008} tunneling
\cite{Zasadzinski:2006,Yazdani:2008} and neutron experiments\ \cite{Fong:2000,
Lipscombe:2007, Vignolle:2007, Wakimoto:2004,Wakimoto:2007}. This work leaves
open the possibility that in the underdoped regime there exists other
instabilities that compete with antiferromagnetism and d-wave superconductivity.

We are grateful to D.J. Scalapino for insights at the origin of this work, to C. Bourbonnais,
J. Carbotte, M. Civelli, T. Maier for discussions, and to M. Greven and G.
Kotliar and L. Taillefer for careful reading and comments on the manuscript.
This work was supported by NSERC (Canada), CFI (Canada), CIFAR, and the Tier I
Canada Research chair Program (A.-M.S.T.). Computations were carried out on
clusters of the R\'{e}seau qu\'{e}b\'{e}cois de calcul de haute performance
(RQCHP) and on the Elix cluster at Universit\'{e} de Sherbrooke. A.-M.S.T.
thanks the Aspen Center for Physics.



\section{Methods, supplement}

In Cellular Dynamical Mean-Field Theory (C-DMFT) a cluster is embedded in a
bath of non-interacting electrons that simulates the effect of the rest of the
infinite lattice by injecting and removing electrons on the cluster with the
appropriate single-particle propagator. The bath is determined
self-consistently by requiring that the self-energy of the infinite system and
that of the cluster be the same. To break the symmetry, frequency independent
source fields are allowed on bath sites only. More details can be found in \cite{Kancharla:2008}.
All the calculations are done with exact
diagonalization at zero temperature \cite{Caffarel:1994}.

For the case of a $2\times2$ plaquette, which we shall consider throughout
this work, the Nambu spinor is defined by
$\Psi_{d}^{\dagger}\equiv(c_{1\uparrow}^{\dagger},\dots,c_{4\uparrow}%
^{\dagger},c_{1\downarrow},\dots,c_{4\downarrow})$, and the greek letters
$\mu,\nu=1,...,N_{c}$ label the degrees of freedom within the cluster. We
compute the cluster propagator $\widehat{G}_{c}$ by solving the cluster
impurity Hamiltonian that will be described shorty. Given the $\hat
{\mathcal{G}}_{0}$ on the cluster that results from the presence of the bath,
we extract the cluster self energy from $\hat{\Sigma}_{c}=\hat{\mathcal{G}%
}_{0}^{-1}-\widehat{G}_{c}^{-1}$. Here,
\begin{equation}
\widehat{G}_{c}\left(  \tau,\tau^{\prime}\right)  =\left(
\begin{array}
[c]{cc}%
\hat{G}_{\uparrow}\left(  \tau,\tau^{\prime}\right)  & \hat{F}\left(
\tau,\tau^{\prime}\right) \\
\hat{F}^{\dagger}(\tau,\tau^{\prime}) & -{\hat{G}}_{\downarrow}\left(
\tau^{\prime},\tau\right)
\end{array}
\right)  \label{nambugreen}%
\end{equation}
is an $8\times8$ matrix, $G_{\mu\nu,\sigma}\equiv-\langle Tc_{\mu\sigma}%
(\tau)c_{\nu\sigma}^{\dagger}(0)\rangle$ and $F_{\mu\nu}\equiv-\langle
Tc_{\mu\uparrow}(\tau)c_{\nu\downarrow}(0)\rangle$ are the imaginary-time
ordered normal and anomalous Green functions respectively.

Using the self-consistency condition,
\begin{equation}
\hat{\mathcal{G}}_{0}^{-1}(i\omega_{n})=\left[  \frac{N_{c}}{(2\pi)^{2}}\int
d\tilde{\mathbf{k}}\;\widehat{G}(\widetilde{\mathbf{k}},i\omega_{n})\right]
^{-1}+\hat{\Sigma}_{c}(i\omega_{n}) \label{selfcon}%
\end{equation}
with
\begin{equation}
\widehat{G}(\widetilde{\mathbf{k}},i\omega_{n})=\left[  i\omega_{n}+\mu
-\hat{t}(\widetilde{\mathbf{k}})-{\hat{\Sigma}_{c}}(i\omega_{n})\right]
^{-1}, \label{Lattice_G}%
\end{equation}
we recompute the Weiss field $\hat{\mathcal{G}}_{0}^{-1},$ obtain the
corresponding bath parameters by minimizing a distance function described
below, and iterate till convergence. Here $\hat{t}(\widetilde{\mathbf{k}})$ is
the Fourier transform of the superlattice hopping matrix with appropriate sign
flip between propagators for up and down spin and the integral over
$\widetilde{\mathbf{k}}$ is performed over the reduced Brillouin zone of the superlattice.

A $2\times2$ plaquette is embedded in a bath of non-interaction electrons. To
solve the cluster impurity problem, we express it in the form of a Hamiltonian
$H_{\mathrm{imp}}$ with a discrete number of bath orbitals coupled to the
cluster and use the exact diagonalization technique (Lanczos method)
\cite{Caffarel:1994} :
\begin{align}
H_{\mathrm{imp}}  &  \equiv\sum_{\mu\nu\sigma}E_{\mu\nu\sigma}c_{\mu\sigma
}^{\dagger}c_{\nu\sigma}+\sum_{m\sigma}\epsilon_{m\sigma}^{\alpha}a_{m\sigma
}^{\dagger\alpha}a_{m\sigma}^{\alpha}\nonumber\\
&  +\sum_{m\mu\sigma}V_{m\mu\sigma}^{\alpha}a_{m\sigma}^{\dagger\alpha}%
(c_{\mu\sigma}+\mathrm{h.c.})+U\sum_{\mu}n_{\mu\uparrow}n_{\mu\downarrow
}\nonumber\\
&  +\sum_{\alpha}\Delta^{\alpha}(a_{1\uparrow}^{\alpha}a_{2\downarrow}%
^{\alpha}-a_{2\uparrow}^{\alpha}a_{3\downarrow}^{\alpha}+a_{3\uparrow}%
^{\alpha}a_{4\downarrow}^{\alpha}-a_{4\uparrow}^{\alpha}a_{1\downarrow
}^{\alpha}\nonumber\\
&  +a_{2\uparrow}^{\alpha}a_{1\downarrow}^{\alpha}-a_{3\uparrow}^{\alpha
}a_{2\downarrow}^{\alpha}+a_{4\uparrow}^{\alpha}a_{3\downarrow}^{\alpha
}-a_{1\uparrow}^{\alpha}a_{4\downarrow}^{\alpha}+h.c.).\nonumber
\end{align}
Here $\mu,\nu=1,...,N_{c}$ label the sites in the cluster and $E_{\mu\nu
\sigma}$ represents the hopping and the chemical potential within the cluster.
The energy levels in the bath are grouped into replicas of the cluster
($N_{c}=4$) (two replicas in the present case) with the labels $m=1,\cdots
,N_{c}$ and $\alpha=1,2$ such that we have $16$ bath energy levels
$\epsilon_{m\sigma}^{\alpha}$ coupled to the cluster via the bath-cluster
hybridization matrix $V_{m\mu\sigma}^{\alpha}$. Using lattice symmetries we
take $V_{m\mu\sigma}^{\alpha}\equiv V^{\alpha}\delta_{m\mu}$ and
$\epsilon_{m\sigma}^{\alpha}\equiv\epsilon^{\alpha}$. The quantity
$\Delta^{\alpha}$ represents the amplitude of superconducting correlations in
the bath. No static mean-field order parameter acts directly on the cluster sites.

The parameters $\epsilon^{\alpha}$, $V^{\alpha}$ and $\Delta^{\alpha}$ are
determined by imposing the self-consistency condition in Eq.~\ref{selfcon}
using a conjugate gradient minimization algorithm with a distance function
\begin{equation}
d=\sum_{\omega_{n},\mu,\nu}\left\vert \left(  \hat{\mathcal{G}}_{0}^{\prime
-1}(i\omega_{n})-\hat{\mathcal{G}}_{0}^{-1}(i\omega_{n})\right)  _{\mu\nu
}\right\vert ^{2} \label{distance}%
\end{equation}
that emphasizes the lowest frequencies of the Weiss field by imposing a sharp
cutoff at $\omega_{c}=1.5$. (Energies are given in units of hopping $t,$ and
we take $\hbar=1$.) The distance function in Eq.(\ref{distance}) is computed
on the imaginary frequency axis (effective inverse temperature, $\beta=50$)
since the Weiss field $\hat{\mathcal{G}}_{0}(i\omega_{n})$ is a smooth
function on that axis.

With the bond superconducting order parameter defined as
\begin{equation}
\psi_{\mu\nu}=\left\langle c_{\mu\uparrow}c_{\nu\downarrow}\right\rangle
\end{equation}
we consider $d$-wave singlet pairing ($\psi\equiv\psi_{12}=-\psi_{23}%
=\psi_{34}=-\psi_{41}$). The average is taken in the ground state of the cluster.

All quantities depending on wave vector, including self-energy, are obtained
from the Green function periodization scheme.

The finite size of the bath in the exact-diagonalization technique is an
additional approximation to the CDMFT scheme. The accuracy of this
approximation can be verified by comparing the CDMFT solution for the one-band
Hubbard model with the solution from the Bethe ansatz~\cite{Capone:2004}. We
have also used this comparison in one dimension as a guideline to fix the
choice of parameters in the distance function ($\omega_{c}=1.5$ and $\beta
=50$). These results in one dimension also compare well with those obtained
using the Hirsch-Fye Quantum Monte Carlo algorithm as an impurity solver where
the bath is not truncated~\cite{KyungQMC:2006}. Further, using finite-size
scaling for these low (but finite) temperature
calculations~\cite{KyungQMC:2006}, it was shown that, at intermediate to
strong coupling, a $2\times2$ cluster in a bath accounts for more than $95\%$
of the correlation effect of the infinite size cluster in the single-particle
spectrum. Because of the finite size of the bath, one also needs to use a
finite linewidht broadening $\eta=0.125$ when plotting the figures.

We can also perform an internal consistency check on the effect of the finite
bath on the accuracy of the calculation. With an infinite bath, convergence
insures that the density inside the cluster is identical to the density
computed from the lattice Green function. In practice, we find that there can
be a difference of $\pm0.02$ between the density estimated from the lattice
and that estimated from the cluster. We display results as a function of
cluster density since benchmarks with the one-dimensional Hubbard model show
that, with a finite bath and the procedure described above, one can reproduce
quite accurately Bethe ansatz results for $n\left(  \mu\right)  $ when the
cluster density is used. Nevertheless, we should adopt a conservative attitude
and keep in mind the error estimate mentioned above.

\section{Supplementary figures and appendices}

\subsection{Technical comments on Fig. 1}

The off-diagonal self-energy $\Sigma_{an}^{\prime\prime}$ is extracted from
the periodized Nambu Green function and plotted with $\eta=0.125.$ The maxima
of the plot are marked with black dots. The imaginary part of the local spin
susceptibility $\chi^{\prime\prime}\left(  \omega\right)  $ is calculated on
the cluster and plotted with the same value of $\eta.$ The dots indicate the
maxima in the limit $\eta=0.$ The various densities evaluated on the cluster
are plotted for the following values of $\left(  \mu,\delta\right)  $ starting
from the normal state $:$ $\left(  0.25,0.37\right)  ,$ $\left(
0.375,0.35\right)  ,$ $\left(  0.5,0.33\right)  ,$ $\left(  0.625,0.31\right)
,$ $\left(  0.75,0.29\right)  ,$ $\left(  0.875,0.26\right)  ,$ $\left(
1.0,0.24\right)  ,$ $\left(  1.125,0.22\right)  ,$ $\left(  1.25,0.20\right)
,$ $\left(  1.375,0.18\right)  ,$ $\left(  1.5,0.16\right)  ,$ $\left(
1.625,0.14\right)  ,$ $\left(  1.75,0.13\right)  ,$ $\left(
1.875,0.11\right)  ,$ $\left(  2.0,0.10\right)  ,$ $\left(  2.125,0.08\right)
,$ $\left(  2.25,0.07\right)  ,$ $\left(  2.375,0.05\right)  ,$ $\left(
2.5,0.04\right)  . $

\subsection{Relation between shift in peak position and single-particle gap}

\begin{figure}[ptb]
\includegraphics[angle=0,width=8cm]{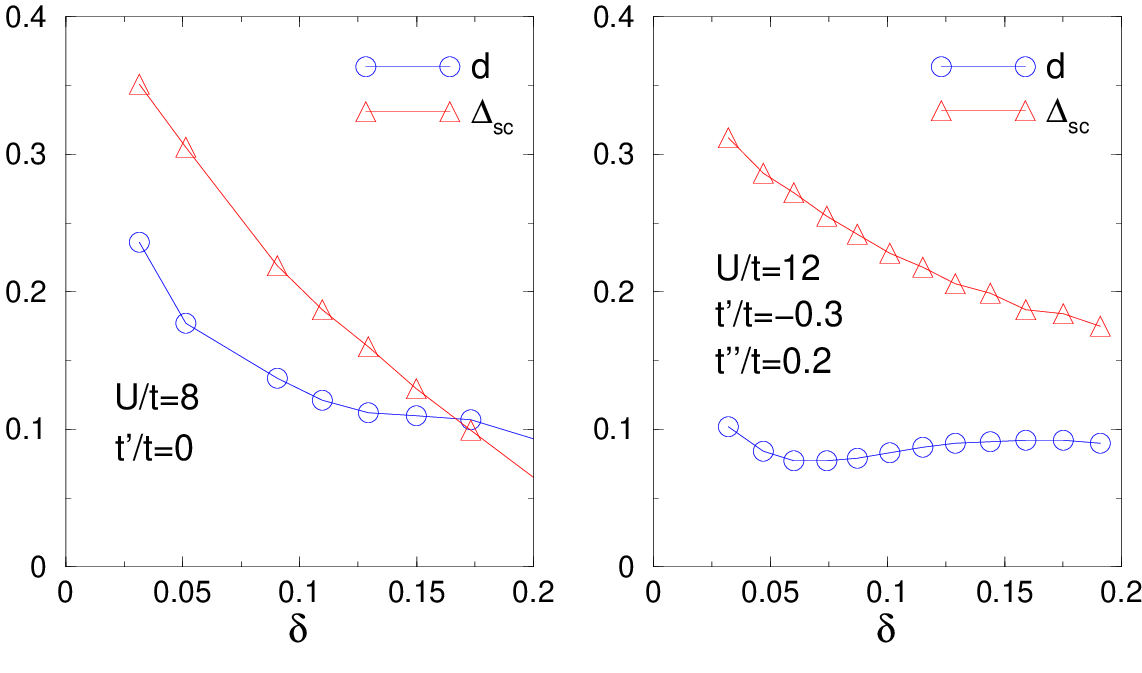}\caption{(Color
online) As a function of doping, the shift $d$ (open circles and blue line)
between the first peak in the $\chi^{\prime\prime}$ and the first peak in
$\Sigma_{an}^{\prime\prime}$ at the antinodal point. Also shown as a function
of doping, the single particle gap $\Delta_{sc}$ (triangles and solid red
line) measured from the single-particle density of states. On the left panel,
$U=8t,$ $t^{\prime}/t=0$. On the right panel $U=12t$ and the band parameters
are those appropriate for YBa$_{2}$Cu$_{3}$O$_{7-\delta}.$ }%
\label{Fig_Delta_vs_Delta}%
\end{figure}

Fig. \ref{Fig_Delta_vs_Delta} illustrates how the single-particle gap in the
superconducting state $\Delta_{sc}$ (not necessarily the superconducting gap)
and the shift $d$ between the position of the peaks in $\Sigma_{an}%
^{\prime\prime}$ and $\chi^{\prime\prime}$ change with doping.

\subsection{Real part of the anomalous self-energy}

\begin{figure}[ptb]
\includegraphics[angle=0,width=8cm]{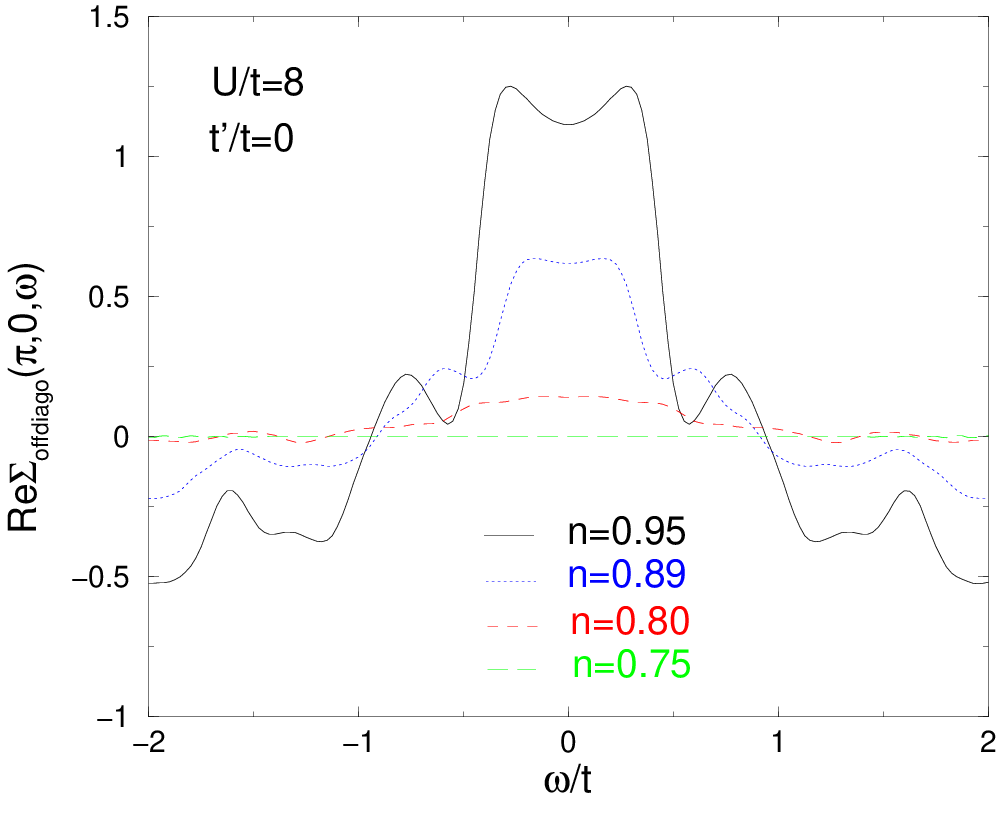}\caption{(Color
online) Real part of the anomalous self-energy $\Sigma_{an}^{\prime}$ for
$U=8,$ $t^{\prime}=t^{\prime\prime}=0$ at the antinodal point. Four different
dopings are presented. Negative contributions appear at a frequency of order
$J$ nearly independent of doping. }%
\label{Fig_Re_Sigma_doping}%
\end{figure}

\begin{figure}[ptb]
\includegraphics[angle=0,width=8cm]{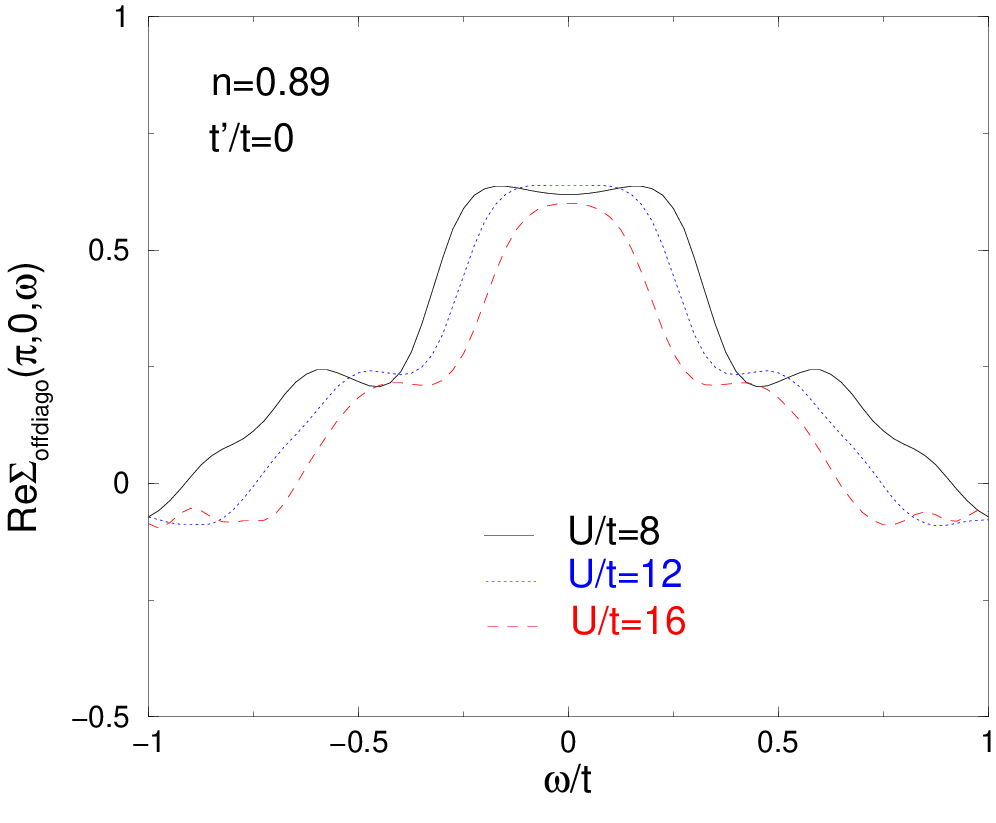}\caption{(Color
online) Real part of the anomalous self-energy $\Sigma_{an}^{\prime}$ as a
function of frequency $\omega$ at the antinodal point for fixed doping
$\delta=0.11$ and different values of $U=8t,12t$ and $16t,$ $t^{\prime
}=t^{\prime\prime}=0$ represented respectively by solid bleue line,
short-dashed red line and long-dashed green line. The nearly flat part near
$\omega=0$ decreases with $J.$ The range of frequencies where $\Sigma
_{an}^{\prime}$ is positive also decreases as $U$ increases or $J$ decreases.
}%
\label{Fig_ReSigma_U}%
\end{figure}

In conventional Migdal-Eliashberg theory, the real part of the self energy
$\Sigma_{an}^{\prime}$ times the quasiparticle renormalization factor is
essentially the gap function. We find that this function, illustrated in Fig
\ref{Fig_Re_Sigma_doping}, increases as one approaches half-filling,
consistent with the increase in the single particle gap found earlier
\cite{Kancharla:2008}
and illustrated in Fig. \ref{Fig_Delta_vs_Delta}.
$\Sigma_{an}^{\prime}$ has weak frequency dependence near zero frequency only
over a range of order $J=4t^{2}/U$ for $U\gtrsim8t,$ as can be seen in Fig.
\ref{Fig_ReSigma_U}. If there were a \textquotedblleft
static\textquotedblright\ piece to the gap, $\Sigma_{an}^{\prime}$ would have
a frequency independent component at frequencies larger than $J$, at least
until frequencies of order $U.$ We find that this is not the case. For the
$t-J$ model \cite{Maier:2008,Haule:2007} 
one finds a small instantaneous
contribution to $\Sigma_{an}^{\prime}$, thus making connection with mean-field
theories. We show in the main text how mean-field theories can also be seen as
approximations to the present approach, even though we do not find an
instantaneous contribution to pairing.

\begin{figure}[ptb]
\includegraphics[angle=0,width=8cm]{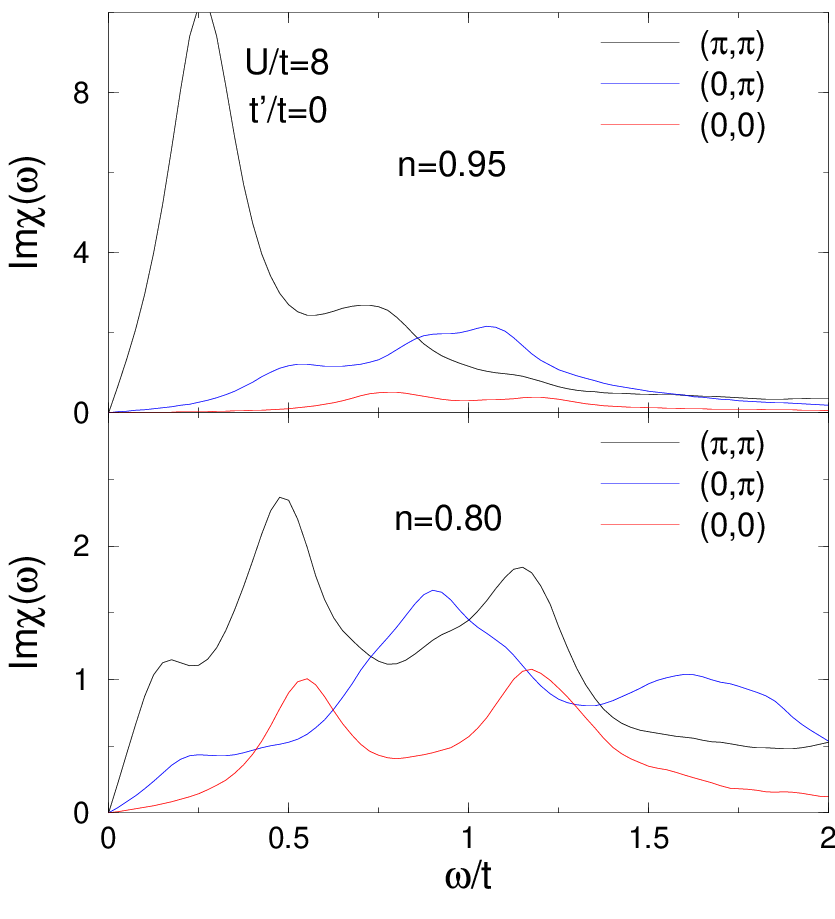}\caption{(Color
online) Imaginary part of the spin susceptibility $\chi^{\prime\prime}$ for
$U=8t,$ $t^{\prime}=t^{\prime\prime}=0$ expressed in cluster momenta for
underdoping on the top panel ($\delta=0.05$) and for overdoping on the bottom
panel ($\delta=0.2$)$.$ Even in the latter case, a sizable $\left(  \pi
,\pi\right)  $ component is left. Recall that the momenta refer to averages
over a quarter of the Brillouin zone.}%
\label{Fig_Spin_Components_Overdoped}%
\end{figure}

\subsection{$I_{G}\left(  \omega\right)  $}

The Lehman representation for the Nambu Green function allows us to find the
following result for the $T=0$ value of $I_{G}\left(  \omega\right)  $%
\begin{equation}
I_{G}\left(  \omega\right)  =\sum_{m}\;\left\langle 0\right\vert c_{i\uparrow
}\left\vert m\right\rangle \left\langle m\right\vert c_{j\downarrow}\left\vert
0\right\rangle \;\theta\left(  \omega-\left(  E_{m}-E_{0}\right)  \right)
\end{equation}
with $\theta$ the Heaviside step function. Excited states $\left\vert
m\right\rangle $ that contribute have an energy less than $\omega$ above the
ground state $\left\vert 0\right\rangle .$

For BCS s-wave theory, $I_{G}\left(  \omega\right)  $ can be computed
analytically. One obtains, using $F_{ij}^{R}$ with $i=j,$
\begin{align}
I_{G}^{BCS}\left(  \omega\right)   &  =\left\langle c_{i\uparrow
}c_{i\downarrow}\right\rangle \bigg[\frac{\sinh^{-1}\left(  \omega
/\Delta\right)  -\sinh^{-1}\left(  1\right)  }{\sinh^{-1}\left(  \omega
_{c}/\Delta\right)  -\sinh^{-1}\left(  1\right)  }\nonumber\\
&  \times\theta\left(  \omega-\Delta\right)  \theta\left(  \omega_{c}%
-\omega\right)  +\theta\left(  \omega-\omega_{c}\right)  \bigg].
\label{Integrated_off_Diagonal_SW}%
\end{align}
The results for the d-wave case in the main text were obtained by numerical
integration and a sharp cutoff.

In Eliashbergh theory, the function $I_{G}$ is%
\begin{equation}
I_{G}\left(  \omega\right)  =N\left(  0\right)  \int_{0}^{\omega
}\operatorname{Re}\left[  \frac{\Sigma_{an}\left(  \omega^{\prime}\right)
}{\sqrt{\left(  Z\left(  \omega^{\prime}\right)  \omega^{\prime}\right)
^{2}-\Sigma_{an}\left(  \omega^{\prime}\right)  ^{2}}}\right]  d\omega
^{\prime}%
\end{equation}
where $N\left(  0\right)  $ is the single-particle density of states at the
Fermi level, the square root is in the upper half-plane and
\begin{equation}
Z\left(  \omega\right)  \equiv1-\frac{\Sigma_{11}\left(  \omega\right)
+\Sigma_{22}\left(  \omega\right)  }{2\omega}%
\end{equation}
with $\Sigma_{ii}\left(  \omega\right)  $ the diagonal components of the
self-energy in Nambu space, $\Sigma_{22}\left(  \omega\right)  =-\Sigma
_{11}\left(  -\omega\right)  $ and $\Sigma_{an}\left(  \omega\right)
\equiv\Sigma_{12}\left(  \omega\right)  $. The phase is chosen such that there
is no contribution from the second Pauli matrix in Nambu space \cite{Scalapino:1966}.

The anomalous Green function $F_{ij}^{R}\left(  \omega\right)  $ entering the
calculation of $I_{G}\left(  \omega\right)  $ in CDMFT was obtained by Fourier
transforming the anomalous lattice Green function calculated with the band
Lanczos approach. We used three different values $\eta=0.24,0.18$ and $0.12$
for the small imaginary part that must be added to the real frequency to
obtain the retarded $F_{ij}^{R}\left(  \omega\right)  .$ The final result for
$I_{G}\left(  \omega\right)  $ is the extrapolation to $\eta=0.$ This is done
to smooth the function while preserving as much as possible the asymptotic
large frequency value. It differs by only a few percent from the value of the
order parameter calculated on the cluster.

\subsection{Effect of the self-energy in decreasing the pairing tendency as
one approaches half-filling}

\begin{figure}[ptb]
\includegraphics[angle=0,width=8cm]{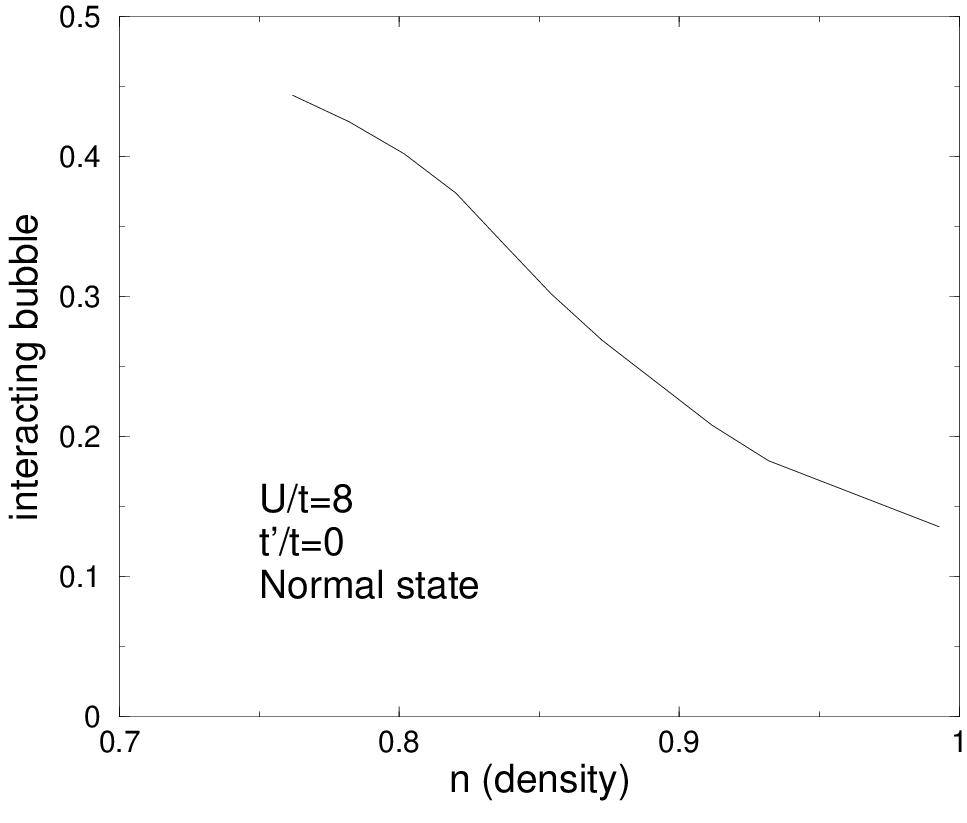}\caption{(Color
online) Pairing susceptibility calculated with the dressed bubble only, i.e.
without vertex corrections. The decrease near half-filling illustrates the
pair-breaking effect of the pseudogap.}%
\label{Pairing_Suscep}%
\end{figure}

Fig. \ref{Pairing_Suscep} shows that in the normal state, the pairing
susceptibility calculated without vertex corrections decreases as one
approaches haff-filling. This is an illustration of the detrimental effect of
the pseudogap. The self-energy in the dressed Green functions entering the
calculation leads to a decrease in the number of states that can pair near the
Fermi level.

\subsection{Attractive Hubbard model}

The cutoff frequency enters very clearly in the integrated off-diagonal
spectral weight Eq.\ (\ref{Integrated_off_Diagonal_SW}). Since one expects
that the attractive (instead of repulsive) Hubbard model should behave more
like the BCS model, we checked that $I_{G}\left(  \omega\right)  $ calculated
with C-DMFT for that model does have the structure of the BCS result for $s-$
wave. In other words, it vanishes below the gap, and increases monotonically
until a sharp cutoff frequency that depends somewhat on $U$ but is of the
order of the bandwidth, as expected from the mean-field solution. There is
some structure in the frequency dependence that is probably caused in part by
the finiteness of the bath used in the calculation, but does not change the
overall trend.



\end{document}